\documentclass[aps,prb,amsmath,twocolumn,superscriptaddress]{revtex4}

\usepackage{amsmath}
\usepackage{amssymb}

\usepackage{graphicx}
\usepackage{bm} 
\usepackage{url}
\def\mean#1{\left< #1 \right>}

\usepackage{color}
\usepackage[percent]{overpic}
\usepackage[export]{adjustbox}
\usepackage{stackengine}

\DeclareMathOperator{\Tr}{Tr}

\begin{document}

\title{
Current-phase relation and flux-dependent thermoelectricity in Andreev interferometers}
\author{Pavel E. Dolgirev}
\affiliation{Skolkovo Institute of Science and Technology, Skolkovo Innovation Center, 3 Nobel St., 143026 Moscow, Russia}
\author{Mikhail S. Kalenkov}
\affiliation{I.E. Tamm Department of Theoretical Physics, P.N. Lebedev Physical Institute, 119991 Moscow, Russia}
\affiliation{Moscow Institute of Physics and Technology, Dolgoprudny, 141700 Moscow region, Russia}

\author{Andrei D. Zaikin}
\affiliation{Institut f{\"u}r Nanotechnologie, Karlsruher Institut f{\"u}r Technologie (KIT), 76021 Karlsruhe, Germany}
\affiliation{P.L. Kapitza Institute for Physical Problems, 119334 Moscow, Russia}

\date{\today}

\begin{abstract}
We predict a novel $(I_0,\phi_0)$-junction state of multi-terminal Andreev interferometers that emerges from an interplay between long-range quantum coherence and non-equilibrium effects. Under non-zero bias $V$ the current-phase relation $I_S(\phi)$ resembles that of a $\phi_0$-junction differing from the latter due to a non-zero average $I_0(V) = \mean{I_S(\phi)}_{\phi}$.  The flux-dependent thermopower ${\mathcal S}(\Phi)$ of the system exhibits features similar to those of a $(I_0,\phi_0)$-junction and in certain limits it can reduce to either odd or even function of $\Phi$ in the agreement with a number of experimental observations.
\end{abstract}

\maketitle

\section{Introduction}
Multi-terminal heterostructures composed of interconnected superconducting (S) and normal (N) terminals (frequently called Andreev interferometers) are known to exhibit
non-trivial behavior provided the quasiparticle distribution function inside the system is driven out of equilibrium. For instance,
it was demonstrated both theoretically \cite{V,WSZ,Yip} and experimentally \cite{Teun} that biasing two N-terminals in a four-terminal NS configuration by an
external voltage $V$ one can control both the magnitude and the phase dependence of the supercurrent flowing between two S-terminals and -- in particular -- provide switching
between zero- and $\pi$-junction states at certain values of $V$. In other words, a $\pi$-junction state in SNS structures can be induced simply by driving
electrons in the N-metal out of equilibrium.

Another way to generate non-equilibrium electron states in Andreev interferometers is to expose the system to a temperature gradient. As a result, an
electric current (and/or voltage) response occurs in the system which is the essence of the thermoelectric effect \cite{Gi}. Usually the magnitude of this
effect in both normal metals and superconductors is small in the ratio between temperature and the Fermi energy $T/\varepsilon_F \ll 1$, however, it can
increase dramatically in the presence of electron-hole asymmetry. The symmetry between electrons and holes in superconducting structures can be lifted for
a number of reasons, such as, e.g., spin-dependent electron scattering (for instance, at magnetic impurities \cite{Kalenkov12}, spin-active interfaces
\cite{KZ15} or superconductor-ferromagnet boundaries \cite{Beckmann}) or Andreev reflection at different NS-interfaces in an SNS structure with a
non-zero phase difference between two superconductors \cite{V2,KZ17} (see also \cite{VH}).
The latter mechanism could be responsible for large thermoelectric signal observed in various types of Andreev interferometers
\cite{Venkat1,Venkat2,Petrashov03,Venkat3,Petrashov16}.

Yet another important feature of some of the above observations is that the detected thermopower was found to oscillate as a function of
the applied magnetic flux $\Phi$ with the period equal to the flux quantum
$\Phi_0=\pi c/e$, thus indicating that the thermoelectric effect essentially depends on the phase of electrons in the interferometer. The symmetry of
such thermopower oscillations was observed to be either odd or even in $\Phi$ depending on the sample topology \cite{Venkat1}. Also, with increasing
bias voltage these oscillations were found to vanish and then re-appear at yet higher voltages with the phase shifted by $\pi$ \cite{Petrashov03}.
Despite subsequent attempts to attribute the results \cite{Venkat1} to charge imbalance effects \cite{Titov2008} or mesoscopic fluctuations
\cite{JW} no unified and consistent explanation for the observations \cite{Venkat1,Venkat2,Petrashov03,Venkat3,Petrashov16} has been offered so far.

\begin{figure}
\centering
\includegraphics[width=0.6\linewidth]{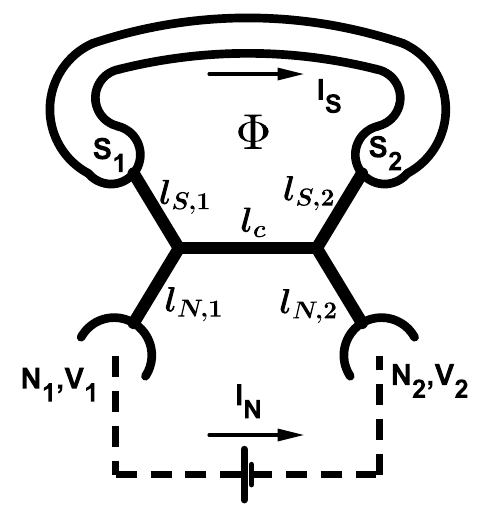}
\caption{A four-terminal structure consisting of five diffusive normal wires of lengths $l_c$, $l_{N,1,2}$ and $l_{S,1,2}$ and cross sections $\mathcal{A}_c$, $\mathcal{A}_{N,1,2}$ and $\mathcal{A}_{S,1,2}$ connecting
two normal terminals biased with a constant voltage $V=V_2 -V_1$ and two superconducting terminals embedded in a
superconducting loop encircling the magnetic flux $\Phi$. We also indicate the currents $I_S$ and $I_N$ flowing respectively in superconducting and normal contours of our setup.}
\label{fig: geom}
\end{figure}

In this paper we address the properties of SNS junctions embedded in multi-terminal configurations with both bias voltage and thermal gradient applied to
different normal terminals. For the configuration depicted in Fig.~\ref{fig: geom} we will demonstrate that at low enough temperatures and with no thermal gradient the corresponding SNS structure
exhibits characteristic features of what we will denote as $(I_0,\phi_0)$-junction state: The current $I_S$ flowing through the superconducting
contour of our setup (as shown in Fig.~\ref{fig: geom}) is predicted to have the form
\begin{equation}
 I_S=I_0(V) + I_1(V, \phi +\phi_0(V)),
\label{Iphi}
\end{equation}
where $I_0 = \mean{I_S}_{\phi}$ and $I_1(V,\phi)$ is a $2\pi$-periodic function of the superconducting phase difference $\phi =2\pi \Phi/\Phi_0$
across our SNS junction. At zero bias
$V \to 0$ both $I_0$ and $\phi_0$ vanish and the term $I_1$ reduces to the equilibrium supercurrent in
diffusive SNS structures \cite{ZZh,GreKa}. At low enough $V$ the contribution $I_1$ essentially coincides with the voltage-controlled Josephson current \cite{WSZ} (with $\phi_0$ jumping from 0 to $\pi$ with increasing $V$), while at higher
voltages with a good accuracy we have $I_1\simeq \tilde I_C(V)\sin (\phi +\phi_0)$ with non-zero phase shift $\phi_0(V)$ which tends to $\pi/2$ in the limit of large $V$. This behavior resembles
that of an equilibrium $\phi_0$-junction which develops nonvanishing supercurrent at $\phi =0$. In contrast to the latter situation,
however, here we drive electrons out of equilibrium, thereby generating extra current $I_0(V)$ along with the phase shift $\phi_0(V)$.
Remarkably, also a thermoelectric signal does not vanish at $\phi =0$ for non-zero $V$, as it will be demonstrated below.

The article is organized as follows. In Section~\ref{sec:QF} we briefly describe the quasiclassical Green function formalism employed in our further analysis. The general current-phase relation for our Andreev interferometer summarized in Eq.~(\ref{Iphi}) is derived and analyzed in Section~\ref{sec: I_0P_0}. In Section~\ref{sec: Thermo} we elaborate on the implications of this relation for the flux-dependent thermopower in multi-terminal Andreev intererometers thereby proposing an interpretation for long-standing experimental puzzles~\cite{Venkat1,Petrashov03}. We close with a brief summary of our key observations in Section~\ref{sec: Conc}.

\section{Quasiclassical formalism}
\label{sec:QF}
In what follows we will employ the quasiclassical Usadel equations which can be written in the
form \cite{belzig1999quasiclassical}
\begin{equation}
iD\nabla (\check G \nabla \check G)=
[\hat1 \otimes \hat\Omega +eV(\bm{r}) , \check G], \quad \check G^2=1.
\label{Usadel}
\end{equation}
Here $4\times 4$ matrix $\check G$ represent
the Green function in the Keldysh-Nambu space
\begin{gather}
\check G=
\begin{pmatrix}
\hat G^R & \hat G^K \\
0 & \hat G^A \\
\end{pmatrix}, \quad
\hat \Omega=
\begin{pmatrix}
\varepsilon & \Delta (\bm{r})\\
-\Delta^* (\bm{r})& -\varepsilon\\
\end{pmatrix},
\end{gather}
$D$ is the diffusion constant, $V(\bm{r})$ is the electric potential, $\varepsilon$ is the quasiparticle energy, $\Delta (\bm{r})$ is
the superconducting order parameter equal to $|\Delta| \exp (i\phi_{1(2)})$ in the first (second) S-terminal and to zero otherwise.
The retarded, advanced, and Keldysh components of the matrix $\check G$ are $2\times 2$ matrices in the Nambu space
\begin{equation}
\hat G^{R,A}=
\begin{pmatrix}
G^{R,A}  & F^{R,A} \\
\tilde F^{R,A} & -G^{R,A} \\
\end{pmatrix}, \quad \hat G^K= \hat G^R \hat f - \hat f \hat G^A,
\end{equation}
where $\hat f = f_L \hat 1+ f_T \hat \tau_3$ is the distribution function matrix
and $\hat\tau_3$ is the Pauli matrix. The current density $\bm{j}$ is related to the matrix $\check G$
by means of the formula
\begin{equation}
\bm{j}= -\frac{\sigma_N}{8 e} \int \Tr [ \hat\tau_3 ( \check G \nabla  \check G)^K]d
\varepsilon ,\label{current}
\end{equation}
where $\sigma_N$ is the Drude conductivity of a normal metal.

Resolving Usadel equations \eqref{Usadel}
for $\hat G^{R,A}$ in each of the normal wires, we evaluate both the spectral current and the
kinetic coefficients \cite{belzig1999quasiclassical}
\begin{gather}
{\bm j}_\varepsilon = \frac{1}{4} \Tr{\hat\tau_3 (\hat{G}^R \nabla \hat{G}^R - \hat{G}^A \nabla \hat{G}^A)},
\label{6}\\
D_L = \frac{1}{2} - \frac{1}{4} \Tr{\hat{G}^R \hat{G}^A},
\label{7}\\
D_T = \frac{1}{2} - \frac{1}{4} \Tr{\hat{G}^R \hat\tau_3 \hat{G}^A \hat\tau_3},
\label{8}\\
\mathcal{Y} = \frac{1}{4} \Tr{\hat{G}^R\hat\tau_3 \hat{G}^A},
\label{9}\end{gather}
which enter the kinetic equations as
\begin{gather}
\nabla {\bm j}_L = 0 ,
\quad
{\bm j}_L =  D_L \nabla f_L - \mathcal{Y} \nabla f_T + {\bm j}_\varepsilon f_T,
\label{eq:j_L}
\\
\nabla {\bm j}_T = 0 ,
\quad
{\bm j}_T =  D_T \nabla f_T + \mathcal{Y}\nabla f_L + {\bm j}_\varepsilon f_L.
\label{eq:j_T}
\end{gather}
Equation (\ref{current}) for the current density can then be cast to the form
\begin{align}
{\bm j} = \frac{\sigma_N}{2e} \int {\bm j}_T d\varepsilon .
\label{cdens}
\end{align}
Analogously one can define the heat current density
\begin{align}
 {\bm j}_Q =  \frac{\sigma_N}{2e^2} \int {\bm j}_L \varepsilon d\varepsilon.
\end{align}

Eqs. \eqref{Usadel} should be supplemented by proper boundary conditions. Here we only address the limit of transparent interfaces and continuously match the normal wires Green functions $\check G$ to those in the normal terminals
\begin{gather}
\hat G^R_{N_i} = - \hat G^A_{N_i} = \hat \tau_3,
\\
f_{L/T,N_i}
=\dfrac{1}{2}
\left[
\tanh \dfrac{\varepsilon + eV_i}{2T_i}
\pm
\tanh \dfrac{\varepsilon - eV_i}{2T_i}
\right],
\label{distrf}
\end{gather}
and in the superconducting ones
\begin{gather}
\hat G^{R,A}
=
\pm
\dfrac{
\begin{pmatrix}
\varepsilon & \Delta \\
- \Delta^* & - \varepsilon
\end{pmatrix}
}{
\sqrt{(\varepsilon\pm i \delta)^2 - \Delta^2}
},
\\
\hat G^K = (\hat G^R - \hat G^A) \tanh\dfrac{\varepsilon}{2 T}.
\end{gather}
The spectral currents ${\bm j}_\varepsilon, {\bm j}_T, {\bm j}_L$ obey the Kirchhoff-like equations in all nodes of our structure.

\section{$(I_0,\phi_0)$-junction}
\label{sec: I_0P_0}
We first consider a symmetric four-terminal setup of Fig.~\ref{fig: geom} with wire lengths
$l_{S(N),1}=l_{S(N),2} = l_{S(N)}$, equal cross sections $\mathcal{A}_{S(N),1}=\mathcal{A}_{S(N),2} = \mathcal{A}_{c}= \mathcal{A}$  and voltages\cite{FN} $V_{1/2} = \mp V/2$. The spectral part of the Usadel equation~(\ref{Usadel}) is solved numerically in a straightforward manner (cf, e.g., Ref. \onlinecite{WSZ}). This solution enables us to find the retarded and advanced Green functions $\hat G^{R,A}$ and to evaluate the spectral current ${\bm j}_\varepsilon$ (\ref{6}) as well as the
kinetic coefficients (\ref{7})-(\ref{9}). In order to resolve the kinetic equations and to determine the current-phase relation for our setup we will adopt the following strategy. We first obtain a simple approximate analytic solution and then verify it by a rigorous numerical analysis.

Let us for a moment assume that the phase difference $\phi$ is small as compared to unity and relax this assumption in the very end of our calculation. In this case one can proceed perturbatively and resolve the kinetic equations in the first order in $j_\varepsilon \propto \phi$. Within the same accuracy, one can drop the small terms $\sim \mathcal{Y}$ and neglect the energy dependence of $D_L \approx 1$. With the aid of Eq. (\ref{cdens}) we arrive at the expressions for the spectral currents $I_{S(N)}(\varepsilon)=\sigma_N j_T\mathcal{A}/(2e)$ flowing in the superconducting (normal) contours of our circuit \cite{FNC}, see Fig.~\ref{fig: geom}. We obtain
\begin{gather}
I_{S}(\varepsilon) = \sigma_Nf^0_L j_
\varepsilon \mathcal{A}/(2e) - f^0_T \mathcal{R}_{c}^T /\mathcal{N}
\label{eq: supercurrent}
\\
I_N(\varepsilon) =  - f^0_T (\mathcal{R}_c^T + 2 \mathcal{R}_S^T)/\mathcal{N}
\label{eq: normcurrent},
\end{gather}
where we defined
\begin{equation}
\mathcal{N}=\mathcal{R}_c^T (\mathcal{R}_S^T + \mathcal{R}_N^T)+
2\mathcal{R}_S^T \mathcal{R}_N^T
\end{equation}
and the spectral resistances $\mathcal{R}_i^T = (\mathcal{A} \sigma_N)^{-1} \int_{l_i} dx/D_{T,i}$
(which reduce to that for a normal wire of length $l_i$ in the normal state with $D_T\equiv 1$).
The distribution functions $f^0_{L/T}$ are given by Eq. \eqref{distrf} with $V_i \to V/2$ and $T_i \to T$.
Integrating Eqs.~(\ref{eq: supercurrent}) and~(\ref{eq: normcurrent}) over energy
$\varepsilon$ we obtain approximate expressions for the currents $I_N$ and $I_N$.

\begin{figure}
\centering
\def\big{\includegraphics[width=1\linewidth, center]{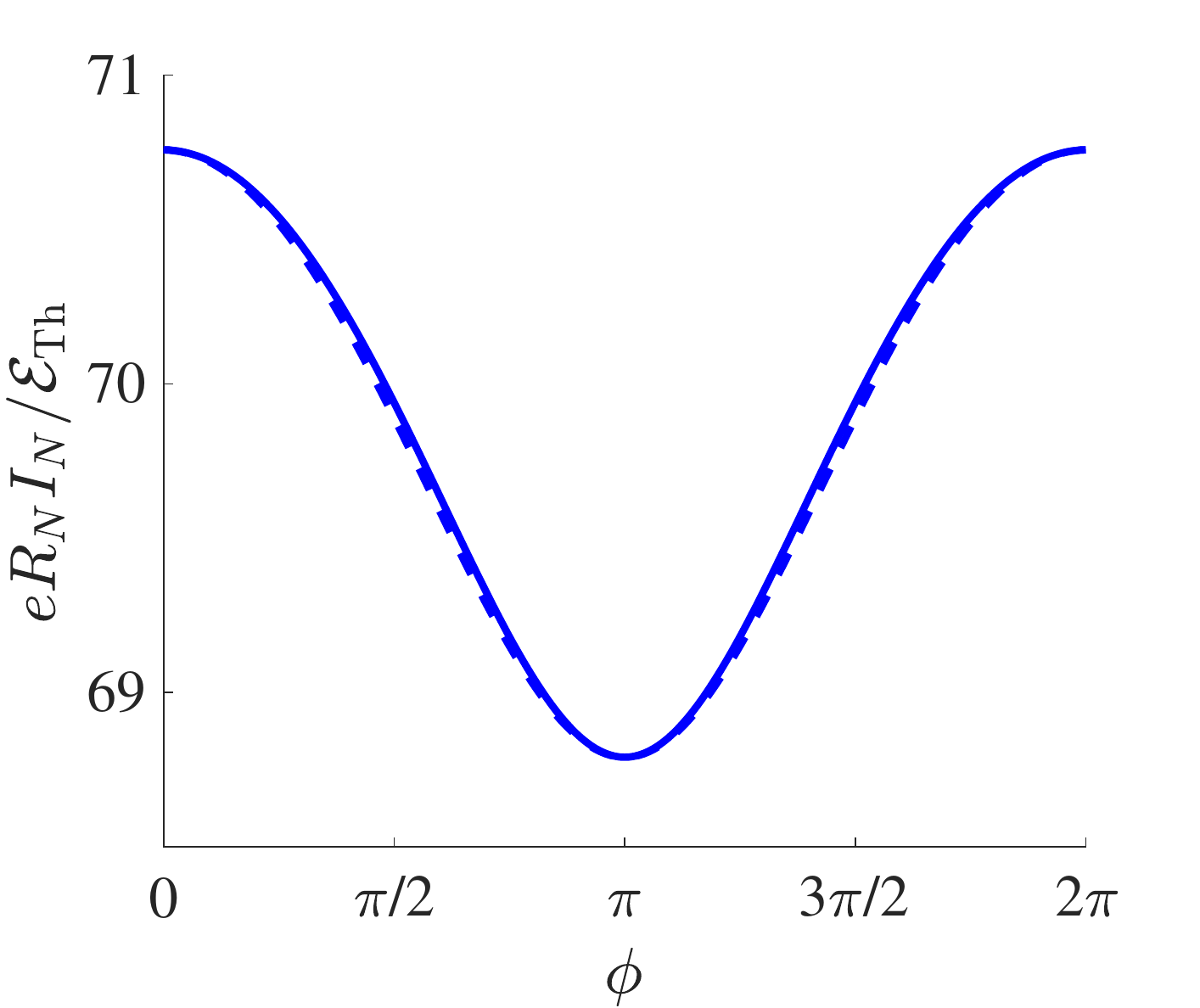}}
\def\little{\includegraphics[height=2.6cm]{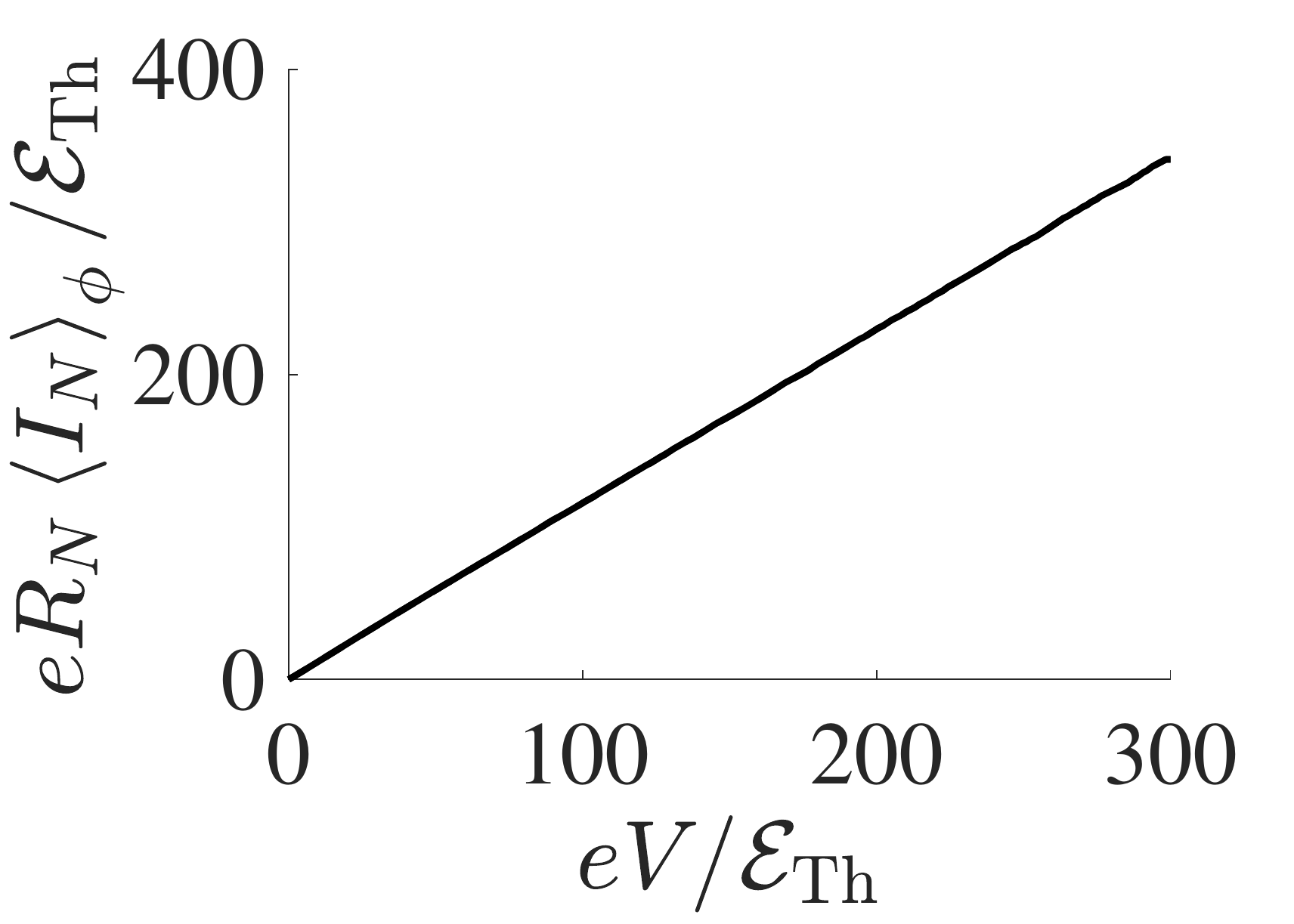}}
\def\stackalignment{l}
\topinset{\little}{\big}{-10pt}{75pt}
\caption{(Color online) The phase dependence of the current $I_N$ at $ T \to 0$ for $eV = 60 \mathcal{E}_{\rm Th}$. Here we set $l_{S}=l_{N}=l_c$ and $\mathcal{E}_{\rm Th} = 10^{-3}\Delta$. The result for $I_N (\phi)$ derived from an approximate Eq.~(\ref{eq: normcurrent}) is indicated by the dashed line, the solid line correspond to our exact numerical solution. The inset illustrates $\mean{I_N}_\phi$ as a function of the voltage bias $V$.}
\label{fig: I_N}
\end{figure}

\begin{figure}
\centering
\def\big{\includegraphics[width=1\linewidth, center]{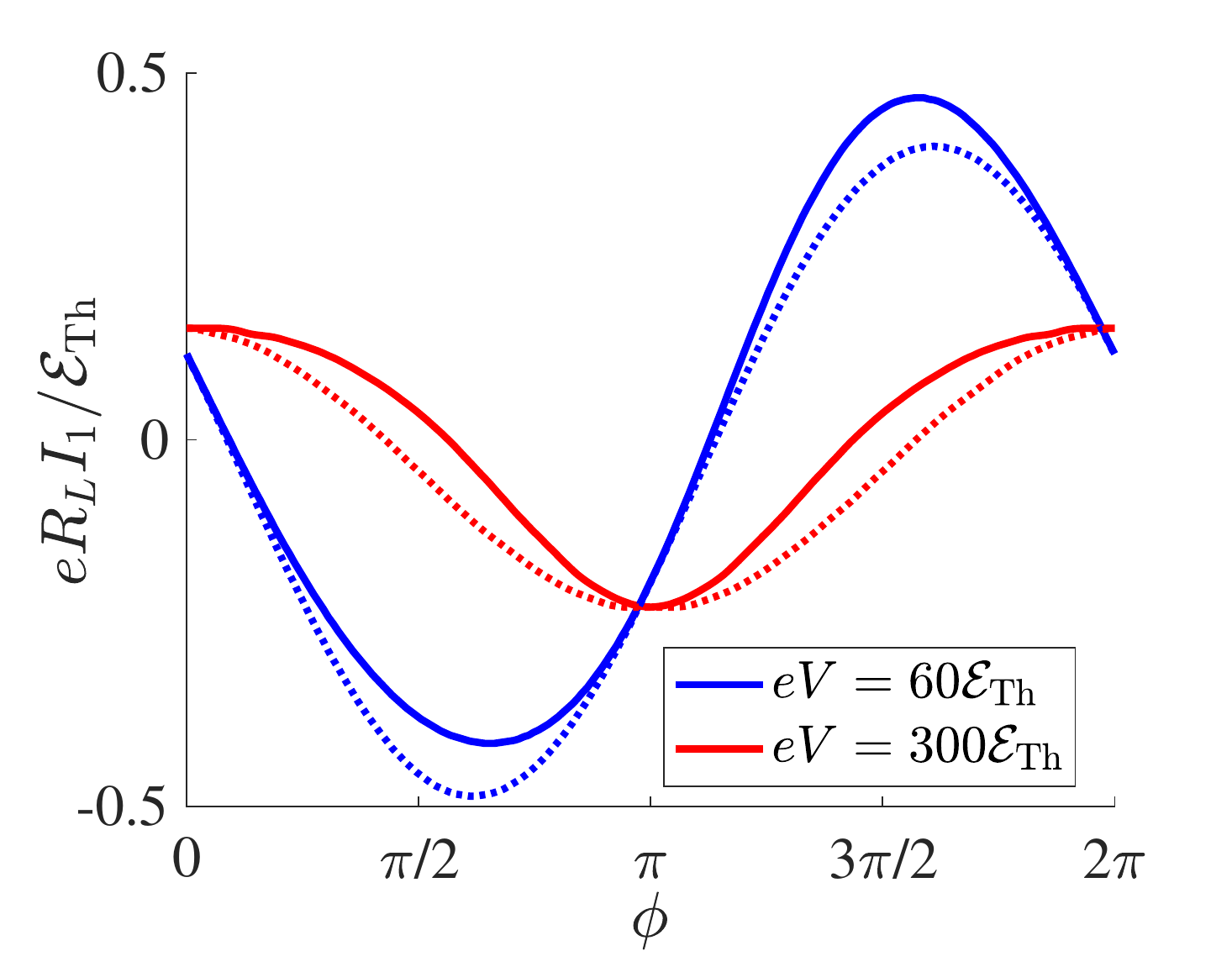}}
\def\little{\includegraphics[height=2.9cm]{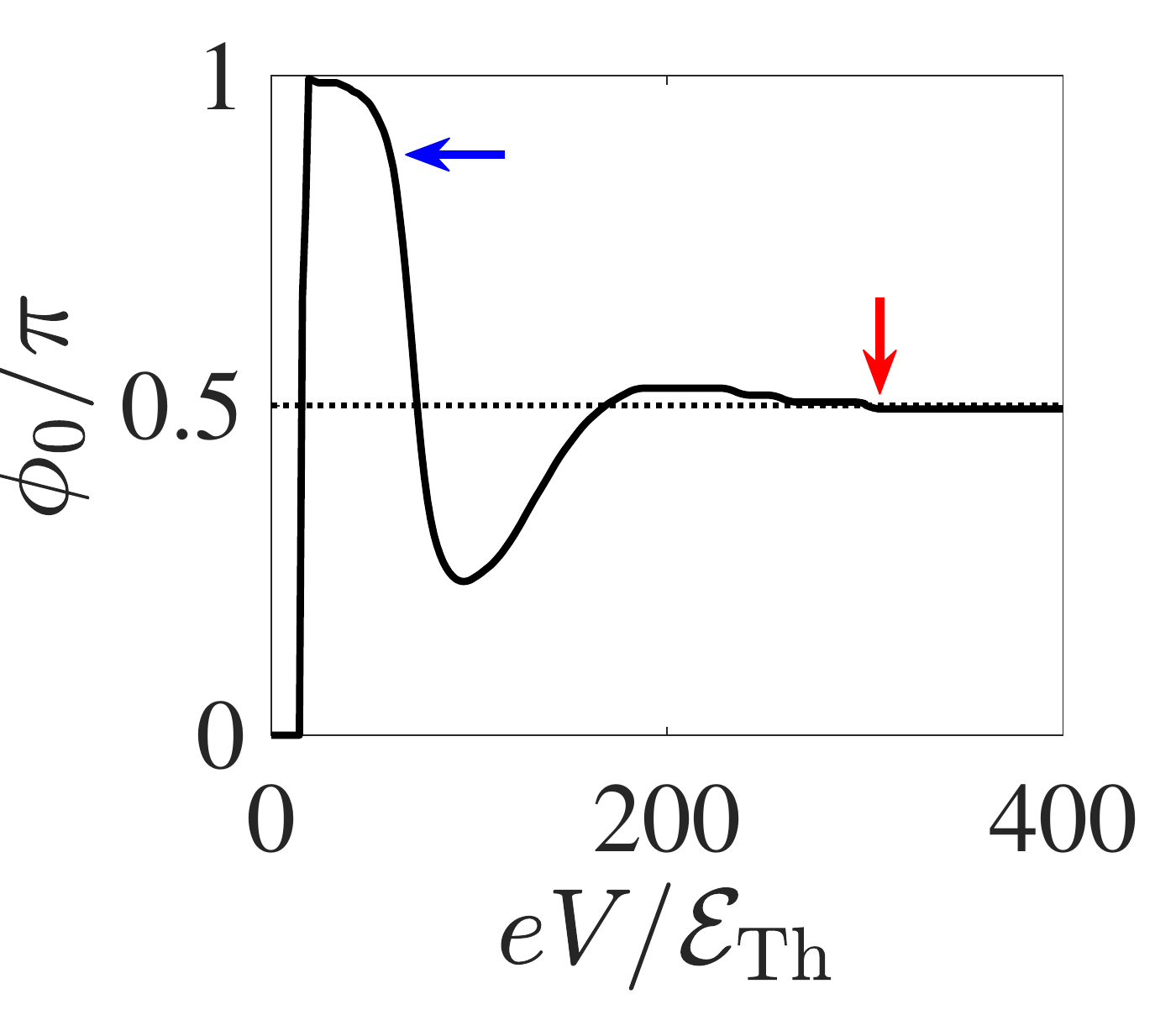}}
\def\stackalignment{l}
\topinset{\little}{\big}{-17pt}{50pt}
\caption{(Color online) The phase dependence of the current $I_1$ at $ T \to 0$ for $eV = 60 \mathcal{E}_{\rm Th}$ and $eV = 300 \mathcal{E}_{\rm Th}$.
Solid lines indicate our exact numerical solution. Dotted lines correspond to a simple analytic expression for $I_1(\phi)$ derived from Eq. \eqref{eq: supercurrent}. The parameters
are the same as in Fig.~\ref{fig: I_N}. Inset: The phase shift $\phi_0$ as a function of $V$. Arrows indicate the voltage values $eV = 60 \mathcal{E}_{\rm Th}$ and $eV = 300 \mathcal{E}_{\rm Th}$.}
\label{fig: MainRes_VirtGeom}
\end{figure}

In addition to the above perturbative analysis we carried out a rigorous numerical calculation of both $I_S$ and $I_N$ involving no approximations. In the low temperature limit $T \to 0$ the
corresponding results are displayed in Figs.~\ref{fig: I_N} and~\ref{fig: MainRes_VirtGeom} along with approximate results derived from Eqs.~\eqref{eq: supercurrent} and~\eqref{eq: normcurrent} in the same limit. It is satisfactory to observe that our simple perturbative procedure yields very accurate result for the current $I_N(\phi)$ not only for small phases but for all values of $\phi$, see Fig.~\ref{fig: I_N}. This current is an even $2\pi$-periodic function of $\phi$ and
$\mean{I_N}_{\phi} \propto V$. Likewise, for the system under consideration we have $I_0=\mean{I_S}_{\phi} \propto V$.

Below in this section we will mainly concentrate on the phase dependence of the current $I_S$. Fig.~\ref{fig: MainRes_VirtGeom} demonstrates that -- in the agreement with our expectations -- our simple analytic result
for $I_S(\phi)$ derived from Eq.~\eqref{eq: supercurrent} is
quantitatively accurate at sufficiently small phase values or, more generally, at all phases $\phi$ in the vicinity of the points $\pi n$. Moreover, even away from these points Eq. \eqref{eq: supercurrent} remains qualitatively correct capturing all essential features obtained within our rigorous numerical analysis. These considerations yield Eq.~\eqref{Iphi} which represents the first key result of our work.

It is instructive to analyze the above expressions in more details.
The first term in the right-hand side of Eq.~\eqref{eq: supercurrent} is a familiar one. In equilibrium it accounts for dc Josephson current~\cite{ZZh,GreKa}, while at non-zero bias $V$ and in the limit $l_c \to 0$
(in which the last term in Eq.~\eqref{eq: supercurrent} vanishes) it reduces to the results~\cite{WSZ,Yip} demonstrating
voltage-controlled $0-\pi$ transitions in SNS junctions. In contrast, the last term in Eq.~\eqref{eq: supercurrent} is a new one being
responsible for both $I_0$ and $\phi_0$ parts. This term is controlled by the combination $D_T(\phi) \nabla f_T$, where $D_T$ is an even
function of $\phi$. Hence, the net current $I_S(\varepsilon )$ is no longer an odd function of $\phi$.

The physics behind this result is transparent. In the presence of a non-zero bias $V$ a dissipative current component, which we will further label as $I_d(V)$,
is induced in the normal wire segments $l_{S,1}$ and $l_{S,2}$. At NS interfaces this current gets converted into extra ($V$-dependent) supercurrent flowing across a superconducting loop.
Since at low temperatures and energies electrons in normal wires attached to a superconductor remain coherent keeping information about the
phase $\phi$, dissipative currents in such wires also become phase (or flux) dependent demonstrating even in $\phi$
Aharonov-Bohm-like (AB) oscillations \cite{nakano1991quasiparticle,SN96,GWZ97,Grenoble}, i.e.
$I_d(V,\phi)=I_0(V)+I_{AB}(V,\phi )$, where $I_0(V) \propto V$.
Combining this contribution to the current $I_S$ with an (odd in $\phi$) Josephson current $I_J(V,\phi)$ we immediately arrive at Eq. \eqref{Iphi}
with $I_1=I_J+I_{AB}$.

The behavior of the phase shift $\phi_0(V)$ displayed in the inset of Fig.~\ref{fig: MainRes_VirtGeom} is the result of
a trade-off  between Josephson and Aharonov-Bohm contributions to $I_1$. At low bias voltages $I_J$ dominates over $I_{AB}$,
and we have $\phi_0 \approx 0$. Increasing the bias to values $eV \sim 20 \mathcal{E}_{\rm Th}$, in full agreement with previous results~\cite{WSZ}
we observe
the transition to the $\pi$-junction state implying the sigh change of $I_J$. Here we defined the Thouless energy $\mathcal{E}_{\rm Th}=D/L^2 \ll \Delta$, where $L = 2l_S + l_c$ is the total length of three wire segments between two S-terminals (see Fig.~\ref{fig: geom}). At even higher bias voltages both terms $I_J$ and $I_{AB}$
eventually become of the same order. For $v=(eV/ 2 \mathcal{E}_{\rm Th})^{1/2} \gg 1$ and at $T \ll \mathcal{E}_{\rm Th}$
we have \cite{WSZ} $I_J =I_C(V)\sin\phi$, where for our geometry
\begin{eqnarray}
&&I_C(V) \simeq \frac{128(1+v^{-1})}{9(3+2\sqrt{2})}\frac{V}{R_L} e^{ -v} \sin (v+v^{-1}).
\end{eqnarray}
We also approximate~\cite{FN1} $I_{AB} \approx I_{\rm m}\cos{\phi}$, where $I_{\rm m} \approx 0.18\mathcal{E}_{\rm Th}/eR_L$ and $R_L$ is the normal resistance of the wire with length $L$. Hence, for $eV \gg \mathcal{E}_{\rm Th} \gg T$ we obtain
\begin{equation}
I_1 \approx \sqrt{I_C^2+I_{\rm m}^2}\sin (\phi+\phi_0), \quad \phi_0(V) =\arctan\frac{I_{\rm m}}{I_C(V)}.
\nonumber
\end{equation}
The function $\phi_0(V)$ (restricted to the interval $0\leq \phi_0\leq\pi$) shows damped oscillations and saturates to the value $\phi =\pi/2$ in the limit of large $V$, as it is also illustrated in the inset of Fig.~\ref{fig: MainRes_VirtGeom}.

At higher $T > \mathcal{E}_{\rm Th}$ the Josephson current decays exponentially with increasing $T$ whereas the Aharonov-Bohm term
shows a much weaker power-law dependence \cite{GWZ97,Grenoble} $I_{AB} \propto 1/T$, thus dominating the expression for $I_1$ and implying that $\phi_0 \simeq \pi/2$
at such values of $T$.

For completeness, we point out that a $(I_0,\phi_0)$-junction state is also realized  in a cross-like geometry with
$l_c=0$ provided we set $l_{S,1}\neq l_{S,2}$ and $l_{N,1}\neq l_{N,2}$ (see, e.g., Fig.~\ref{fig: Thermoelectric} below).
Under these conditions the distribution function $f_T$ at the wire crossing point differs from zero resulting in a non-vanishing even in $\phi$ contribution to $I_S$ containing $D_T(\phi) \nabla f_T$. However, if either $l_{S,1}= l_{S,2}$ or $l_{N,1}= l_{N,2}$ this even in $\phi$ contribution vanishes and we get back to the results \cite{WSZ,Yip} describing 0- and $\pi$-junction states.

\begin{figure}
\centering
\def\big{\includegraphics[width=1\linewidth]{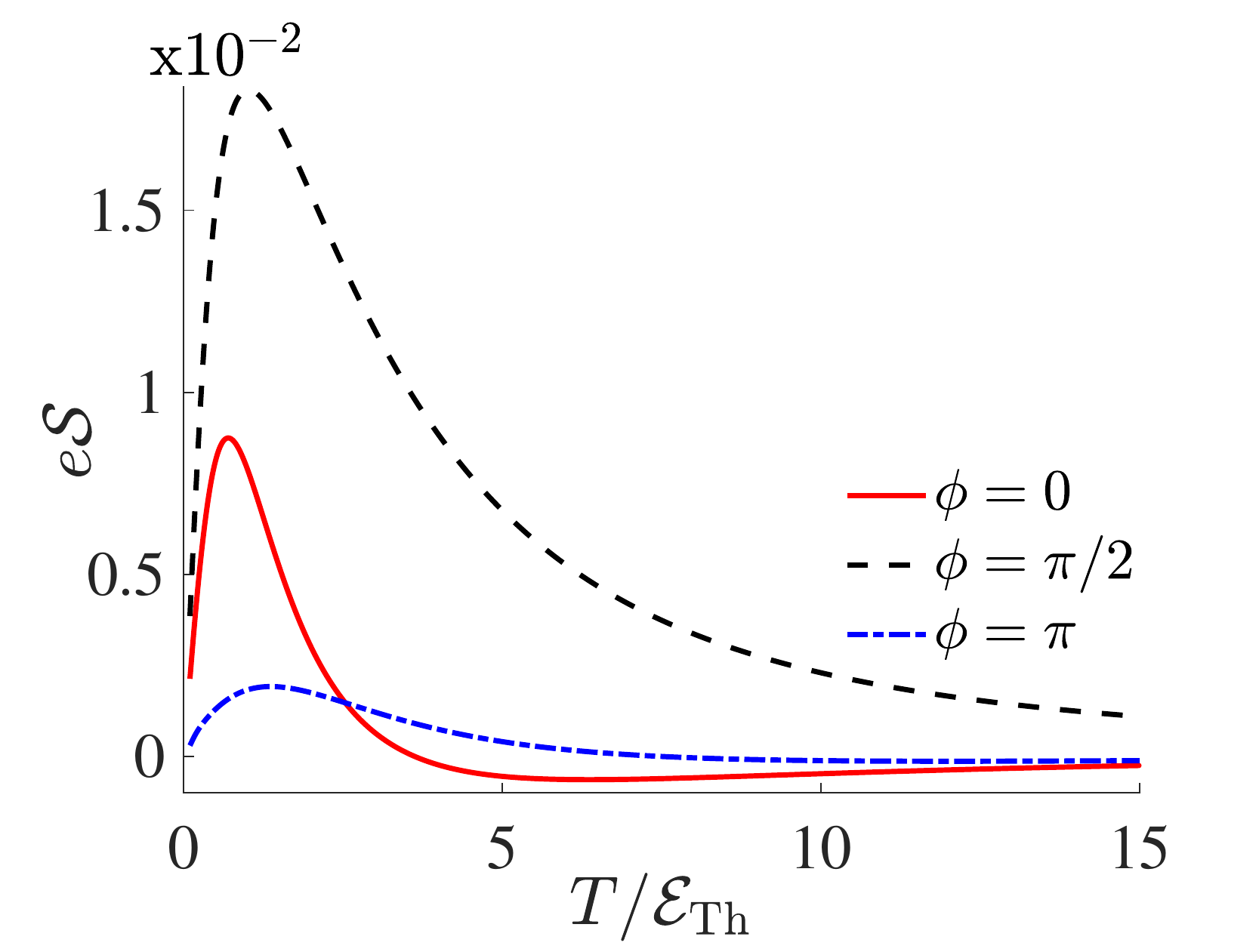}}
\def\little{\includegraphics[height=3.4cm]{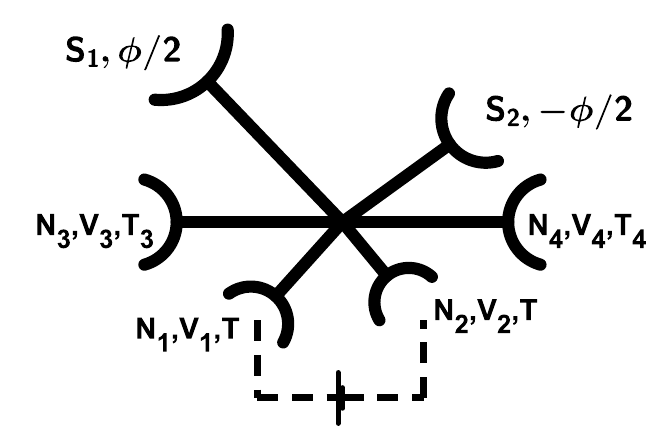}}
\def\stackalignment{r}
\topinset{\little}{\big}{-20pt}{-5pt}
\caption{(Color online) The temperature dependence of the thermopower ${\mathcal S}= V_T/\delta T$ between the terminals N$_3$ and N$_4$ of the six-terminal setup schematically illustrated in the inset.
Different curves correspond to different values of $\phi$. Here we set $eV = 0.9\Delta$,  $\mathcal{E}_{\rm Th} = 10^{-2} \Delta$
and fix the wire lengths as $l_{S,1} = 0.2L, l_{S,2} = 0.8L, l_{N,1} = 0.3L, l_{N,2} = 0.7L$ and $l_{N,3} = l_{N,4} = 0.5L$.}
\label{fig: Thermoelectric}
\end{figure}

\section{Flux-dependent thermopower}
\label{sec: Thermo}
We now turn to the thermoelectric effect. It was argued \cite{V2,KZ17,VH} that
in Andreev interferometers this effect may become large provided the phase difference $\phi$ between superconducting electrodes differs from $\pi n$. Below we will demonstrate that a large thermopower can be induced by a temperature
gradient even if $\phi =0$.

To this end let us somewhat modify the setup in Fig.~\ref{fig: geom} by setting $l_c=0$ and attaching two extra normal terminals N$_3$ and N$_4$ as shown in Fig.~\ref{fig: Thermoelectric}. These terminals are disconnected from the external circuit and are maintained at different temperatures $T_3$ and $T_4$, while the temperature of the remaining four terminals equals to $T$.

We first set $\phi =0$ and evaluate the thermoelectric voltage
$V_{T} = V_3 - V_4$ between N$_3$ and N$_4$ induced by a thermal gradient $\delta T= T_3-T_4$.
For simplicity, below we consider the configuration with $l_{N_3}=l_{N_4}$. As no current can flow into the terminals N$_3$ and N$_4$, we obtain
\begin{equation}
\int \mathcal{G}^T_N(\varepsilon)
\left[ f_{T,N_3} - f_{T,N_4} \right]d \varepsilon =0,
\label{rel}
\end{equation}
where $\mathcal{G}^T_N = 1 / \mathcal{R}^T_{N_3} = 1 / \mathcal{R}^T_{N_4}$ is the spectral conductance. Eq.~(\ref{rel}) defines the relation between $T_3,\ T_4$ and the induced voltages $V_3,\ V_4$. In the first order in $\delta T/T$ it yields the thermoelectric voltage in the form
\begin{equation}
eV_T
=
\dfrac{\delta T }{T}
\dfrac{
\displaystyle\int
\frac{(\varepsilon + eV_N)\mathcal{G}^T_N(\varepsilon) d \varepsilon}{
\cosh^2[(\varepsilon + eV_N)/(2T)]}
}{
\displaystyle\int
\frac{\mathcal{G}^T_N(\varepsilon) d \varepsilon}{
\cosh^2[(\varepsilon + eV_N)/(2T)]}}.
\label{VT}
\end{equation}
Here $V_N(V)$ is the induced electric potential of the terminals N$_3$ and N$_4$ evaluated at $\delta T=0$. For any nonzero bias $V$ the voltage
$V_N$ differs from zero as long as $l_{N,1} \neq l_{N,2}$. In this case the thermovoltage $V_T$ \eqref{VT} also remains nonzero as the
spectral conductance $\mathcal{G}^T_N$ explicitly depends on energy $\varepsilon$ due to the superconducting proximity effect.
On the other hand, in the absence of superconductivity the latter dependence disappears and the expression \eqref{VT} vanishes
identically even for nonzero $V_N$. This observation emphasizes a non-trivial interplay between superconductivity, quantum coherence and thermoelectricity in hybrid metallic nanostructures.

\begin{figure}
\centering
\includegraphics[width=1\linewidth]{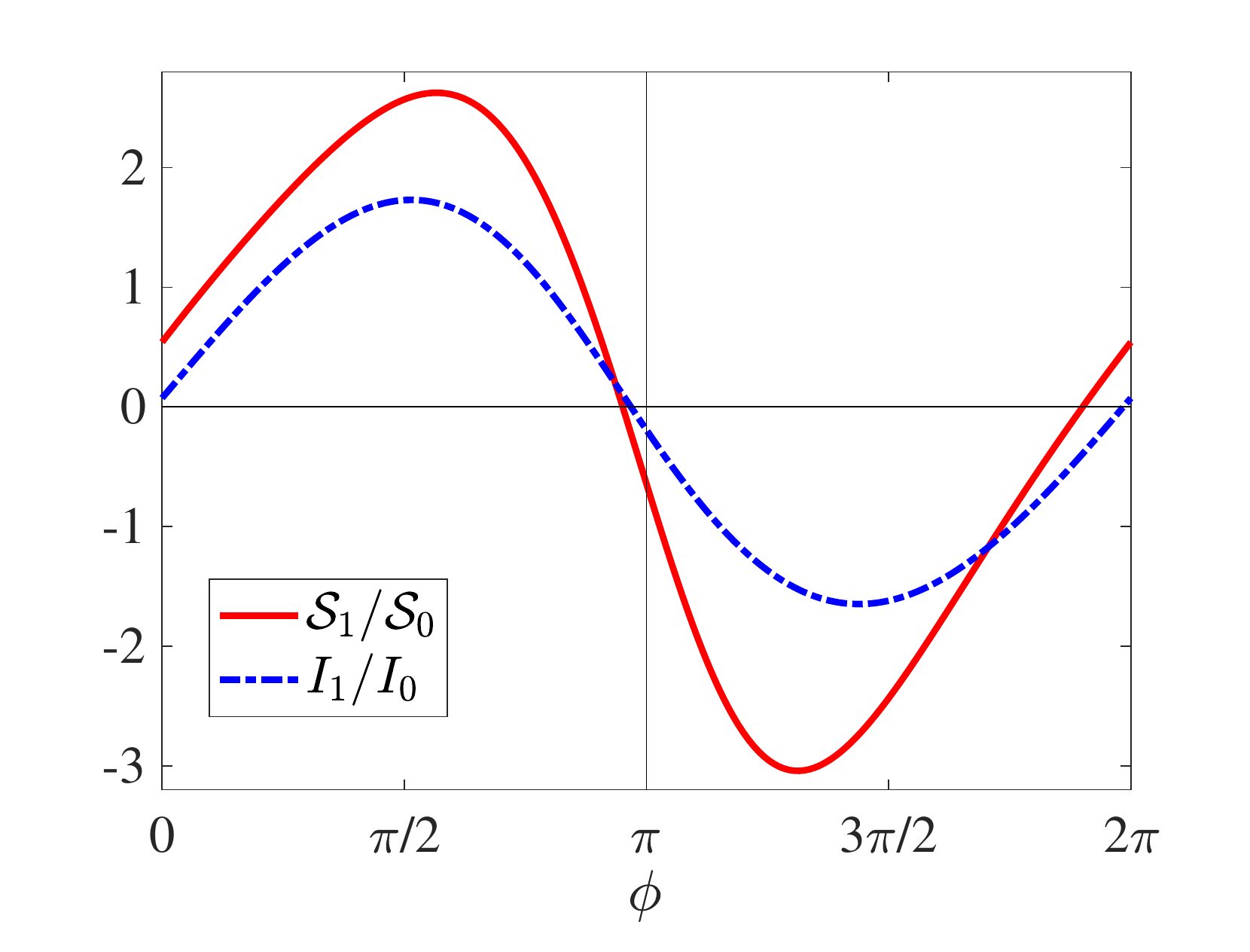}
\caption{(Color online) Thermopower ${\mathcal S}$ and current $I_S$ as functions of $\phi$ evaluated numerically for
the six-terminal setup of Fig.~\ref{fig: Thermoelectric} at $T = \mathcal{E}_{\rm Th}$. The parameters are the same as in Fig. ~\ref{fig: Thermoelectric}.}
\label{fig: ThermoPhi}
\end{figure}

In order to recover the phase dependence of the thermoelectric voltage we treated the problem numerically. The corresponding results are displayed in Figs.~\ref{fig: Thermoelectric} and~\ref{fig: ThermoPhi}. Fig.~\ref{fig: Thermoelectric} demonstrates the temperature dependence of the thermopower ${\mathcal S} = V_T/\delta T$ at different values of $\phi$. In Fig.~\ref{fig: ThermoPhi} we present the thermopower as a function of $\phi$ at $T = \mathcal{E}_{\rm Th}$ together with the current-phase relation $I_S(\phi )$ evaluated for the same setup. We observe that both functions ${\mathcal S}(\phi)$ and $I_S(\phi )$ demonstrate essentially the same behavior and, hence, in complete analogy with Eq.~\eqref{Iphi} we have
\begin{equation}
{\mathcal S} = {\mathcal S}_0(V)+{\mathcal S}_1(V, \phi +\phi'_0(V)),
\label{Vphi}
\end{equation}
where ${\mathcal S}_0=\langle {\mathcal S}(\phi)\rangle_\phi$ and ${\mathcal S}_1(V,\phi)$ is a $2\pi$-periodic function of $\phi$, which at high enough voltages only slightly deviates from a simple form ${\mathcal S}_1(V,\phi) \propto \sin \phi$
(cf. Fig.~\ref{fig: ThermoPhi}).

Eqs.~\eqref{VT},~\eqref{Vphi} represent the second key result of our work. It allows to conclude that in general the periodic dependence of the
thermopower ${\mathcal S}$ on the magnetic flux in Andreev interferometers is neither even nor odd in $\Phi$, but it can reduce to either one of
them depending on the system topology or, more specifically, on the relation between $eV$, $T$ and the relevant Thouless energy $\mathcal{E}_{\rm Th}$.
The phase shift $\phi'_0(V)$ in Eq.~\eqref{Vphi} is not strictly identical to $\phi_0(V)$ in Eq.~\eqref{Iphi}~\cite{FN2}, however, both
these functions behave similarly. In fact, $\phi'_0$ only slightly deviates from $\phi_0$ (cf., e.g, Fig.~\ref{fig: ThermoPhi}). With increasing
$V$, the phase $\phi'_0$ also experiences an abrupt transition from 0 to $\pi$ and then tends to $\pi /2$ in the limit of large voltages and/or temperatures.

Our findings allow to naturally interpret the experimental results \cite{Venkat1} where both odd and even dependencies
of $V_T$ on $\Phi$ were detected depending on the system topology. Indeed, while at small enough $eV$ and $T$ we have
$\phi'_0 \approx 0$ and ${\mathcal S} (\phi)$ remains an odd function, at larger voltages $eV \gtrsim 200 \mathcal{E}_{\rm Th}$
and/or temperatures $T \gg \mathcal{E}_{\rm Th}$ the phase shift approaches $\phi'_0 \simeq \pi/2$ and the flux dependence
of the thermopower ${\mathcal S} (\phi)$ \eqref{Vphi} turns even, just as it was observed  for some of the structures \cite{Venkat1}.
Furthermore, as we already discussed, with increasing bias $V$ the phase $\phi'_0$ jumps from 0 to $\pi$ which is
fully consistent with the observations \cite{Petrashov03}. Thus, we believe the $0-\pi$ transition for the flux-dependent thermopower ${\mathcal S} (\phi)$ detected in experiments \cite{Petrashov03} has the same physical origin as that predicted \cite{V,WSZ,Yip} and observed \cite{Teun} earlier for dc Josephson current.

\section{Summary}
\label{sec: Conc}

In this work we have elucidated a non-trivial interplay between proximity-induced quantum coherence and non-equilibrium effects in multi-terminal hybrid
normal-superconducting nanostructures. We have demonstrated that applying an external bias one drives the system to a
$(I_0,\phi_0)$-junction state in Eq.~(\ref{Iphi}) determined by a trade-off between non-equilibrium Josephson and Aharonov-Bohm-like contributions.  We have
also analyzed the phase-coherent thermopower in such nanostructures which exhibits periodic dependence on the magnetic flux
being in general neither even nor odd in $\Phi$. Our results allow to formulate a clear physical picture explaining a number of
existing experimental observations and calling for further experimental analysis of the issue.

\vspace{0.5cm}

\centerline{\bf Acknowledgements}
We would like to thank A.G. Semenov for fruitful discussions. This work is a part of joint Russian-Greek Projects
No. RFMEFI61717X0001 and No. T4$\Delta$P$\Omega$-00031 Experimental
and theoretical studies of physical properties of low-dimensional
quantum nanoelectronic systems. One of us (P.E.D.) also acknowledges support by Skoltech as a part of Skoltech NGP program and the hospitality of KIT during November 2017.


\begin{thebibliography}{99}
\bibitem{FNC} As it is clear from Fig.~\ref{fig: geom}, the superconducting contour (carrying the current $I_S$) includes the superconductor as well as the normal-wire segments $l_{S,1}$ and $l_{S,2}$, while the normal contour (with the current $I_N$) consists of the battery and the normal wire segments $l_{N,1}$ and $l_{N,2}$. The current through the normal-wire segment $l_c$ obviously equals to $I_S+I_N$.
\bibitem{FN1} Strictly speaking, the $2\pi$-periodic even in $\phi$ function $I_{AB}(\phi)$ slightly deviates
from a simple $\cos\phi$-form, however, a small admixture of higher harmonics $\propto \cos (n\phi)$ with $n \geq 2$ does not alter any of our conclusions.
\bibitem{FN2} According to~\cite{VH}, the thermopower ${\mathcal S}$ is determined by the derivative $dI_S/dT$. Hence, deviations between the phase shifts $\phi_0$ and $\phi'_0$ can be attributed to
different temperature dependencies of Aharonov-Bohm $I_{AB}$ and Josephson $I_{J}$ contributions to $I_1$. For completeness,
we note that the derivative $dI_N/dT$ in general also contributes to the thermopower.
\end{thebibliography}
\end{document}